\documentclass[prl,twocolumn,amsmath,amssymb,superscriptaddress]{revtex4}

\usepackage{graphicx}
\usepackage{dcolumn}
\usepackage{bm}
\usepackage{times}
\newcommand{\bra}[1]{\ensuremath{\langle#1|}}
\newcommand{\ket}[1]{\ensuremath{|#1\rangle}}
\newcommand{\bbra}[1]{\ensuremath{\big\langle\!\!\big\langle#1\big|}}
\newcommand{\kket}[1]{\ensuremath{{\boldmath \big|}#1\big\rangle\!\!\big\rangle}}
\newcommand{\bbrakket}[2]{\ensuremath{\big\langle\!\!\big\langle#1\big|#2\big\rangle\!\!\big\rangle}}
\newcommand{\Tr}{\mbox{Tr}}

\begin{document}
\preprint{}
\title{
Exciton Lifetime Paradoxically Enhanced by Dissipation and Decoherence\\
-- Toward Efficient Energy Conversion of Solar Cell}

\author{Yasuhiro Yamada}
\affiliation{Department of Applied Physics, The University of Tokyo, Hongo, Bunkyo-ku, Tokyo, 113-8656, Japan}
\author{Youhei Yamaji}
\affiliation{Quantum-Phase Electronics Center (QPEC), The University of Tokyo, Hongo, Bunkyo-ku, Tokyo, 113-8656, Japan}
\author{Masatoshi Imada}
\affiliation{Department of Applied Physics, The University of Tokyo, Hongo, Bunkyo-ku, Tokyo, 113-8656, Japan}
\date{\today}

\begin{abstract}
Energy dissipation and decoherence are at first glance harmful to acquiring long exciton lifetime desired for efficient photovoltaics. In the presence of both optically forbidden (namely, dark) and allowed (bright) excitons, however, they can be instrumental as suggested in photosynthesis. By simulating quantum dynamics of exciton relaxations, we show that the optimized decoherence that imposes a quantum-to-classical crossover with the dissipation realizes a dramatically longer lifetime. In an example of carbon nanotube, the exciton lifetime increases by nearly two orders of magnitude when the crossover triggers stable high population in the dark exciton.
\end{abstract}

\pacs{42.50.Ct, 78.67.-n, 84.60.-h, 78.67.Ch}

\maketitle
Sunlight is a clean, abundant, and sustainable energy source. Hence, the effective energy conversion from sunlight into electricity is a grand challenge in science and technology, which leads to an emerging interest on the post-silicon photovoltaic materials such as carbon nanotubes, quantum dots, transition metal dichalcogenides for their promising applications~\cite{Beard:2013bp,Jariwala:2013br,Wang:2012fa}. The energy conversion consists of three processes of nonequilibrium quantum dynamics of excitons (bound electron-hole pairs): Exciton generation from a photon, exciton energy transfer, and charge separation of the exciton into electrodes. In the first two processes, the efficient conversion requires both a high photon absorption rate \textit{and long exciton lifetime}. However, the optimization is hampered by the reversibility between absorption and emission of a photon in the quantum dynamics: While a high absorption rate of photon is desired in the exciton generation process, it also leads to a high charge recombination rate of exciton.

In this work, we show that a desired remarkable enhancement of the exciton lifetime by simultaneously keeping high absorption rate is achieved by utilizing nonequilibrium energy dissipation and decoherence by phonons. Though the dissipation is in general an obstacle to high efficiency of energy conversion, there are cases where nonequilibrium dissipations and decoherence are actually instrumental in achieving long exciton lifetime because they can suppress the exciton recombination by making quantum dynamics irreversible via concomitant quantum-to-classical crossover~\cite{Zurek:2003de}.

In fact, irreversible exciton dynamics is exploited in photosynthesis that also includes the above mentioned three processes, namely, the exciton generation, the exciton energy transfer, and the charge separation~\cite{Scholes:2011lf,Blankenship:2011cp}. Photosynthesis achieves a remarkably high quantum efficiency reaching nearly 100\%~\cite{Wraight:1974gd}, which means that an absorbed photon is converted to an exciton with no recombination. In the exciton generation process, the absorbed energy is irreversibly transferred from the optically allowed bright exciton to an optically forbidden dark exciton~\cite{McDermott:1995cs,Sumi:1999dt,Scholes:2000hu}, which can act as a ratchet between the exciton generation process and the next energy transfer process where the quantum coherence plays a role again~\cite{Engel:2007hb,Rebentrost:2009hu,Plenio:2008ff,Ishizaki:2009ut,Collini:2010fy}.

Recently, photovoltaic models inspired by photosynthesis were studied from the viewpoint of steady-state heat engines modified with discrete exciton states~\cite{Dorfman:2013fm,Creatore:2013ik}. A mechanism for the enhancement of the photocurrent was proposed by designing the ultrafast classical transition from bright to dark excitons within a hundred femtoseconds, where quantum coherence between photons and excitons and the resultant photoluminescence causing the recombination were also ignored~\cite{Creatore:2013ik}. However, these classical descriptions of the photocell dynamics make the validity of the enhancement questionable because the inherent dichotomy between quantum coherence and decoherence is a crucial issue in enhancing the quantum efficiency in the target time domain: On the one hand the desired ultrafast transition from the bright to the dark excitons necessarily requires quantum coherent dynamics, whereas on the other hand, the decoherence that imposes a quantum-to-classical crossover is assumed to immediately occur.

For the microscopic understanding of optimal energy conversions, here we study the transient dynamics for the exciton generation process, taking account of the dichotomy in a unified quantum manner. To this end, we construct and investigate an open quantum model that consists of bright and dark excitons coupled with phonons and a dissipative photon, with using the combined method of the generalized quantum master equation (GQME)~\cite{Petruccione:378945} and the quantum continuous measurement theory of photon counting~\cite{Srinivas:1981hi,Srinivas:1982el}. We found that realistic coexistence of coherence and decoherence is the key for high quantum efficiency of photovoltaics. The crossover from quantum to classical dynamics due to dissipation and decoherence by phonons assists the long exciton lifetime: Rapid quantum energy transfer from the bright to dark excitons suppresses the initial radiative loss, whereas the dark excitons become stable through the concomitant crossover. Our results indeed reveal why high efficiency of exciton generation from photons can be compatible with the low photoluminescence by violating the reversibility as observed experimentally.

\begin{figure}[thbp]
\centering
\includegraphics[keepaspectratio,width=8.6cm]{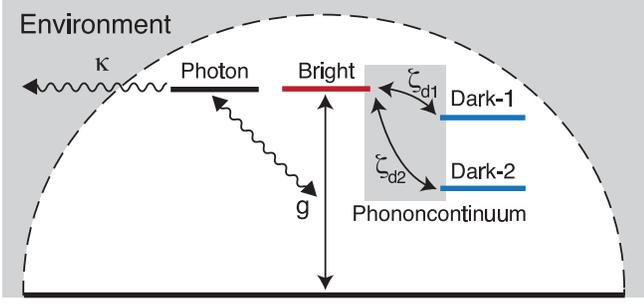}
\caption{(Color online) Dissipative quantum model of excitons in semiconducting single-walled carbon nanotube (SWCNT) with the (6,5) chirality. Both the bright exciton generation and the charge recombination result from the dipole coupling with photons with the same strength $g$. The photons are dissipated to environment with the rate of $\kappa$. The bright exciton is coupled with two dark excitons via a phononcontinuum that causes energy dissipation and decoherence.}
\label{fig1}
\end{figure}

By taking a typical semiconducting single-walled carbon nanotube (SWCNT) with the chirality (6,5), as a model material, we demonstrate that the exciton lifetime becomes nearly two-orders of magnitude longer than the case without dark excitons. Moreover, our results are consistent with the experimental indications~\cite{Berciaud:2008ld,Duque:2009gs,Park:2011jf,Stich:2014kr}. In the SWCNT, at least two dark exciton states exist with energy below the lowest bright one~\cite{Zhao:2004ee,Spataru:2005ta,Ando:2006eo,Tretiak:2007ts,Mortimer:2007ro,Kiowski:2007eb,Shaver:2007mb,Berciaud:2008ld,Matsunaga:2008ef,Shaver:2008mo,Duque:2009gs,Matsunaga:2010ex,Nagatsu:2010bo,Park:2011jf,Konabe:2011bn,Arnold:2013rd,Stich:2014kr}: Even-parity dark exciton~\cite{Matsunaga:2008ef,Shaver:2008mo,Duque:2009gs} and spin-triplet dark exciton~\cite{Matsunaga:2010ex,Nagatsu:2010bo,Stich:2014kr}. The weak couplings between the bright and the dark excitons can strongly affect the exciton dynamics after the photon absorption. Hence, we construct a Hamiltonian consisting of one bright and two dark exciton states coupled to a phononcontinuum and a single photon, as a minimal model of the SWCNT (see also Fig.~\ref{fig1}). The model Hamiltonian is given by
\begin{align}
\hat{H}=\hat{H}_\mathrm{S}+\hat{H}_\mathrm{B}+\hat{H}_\mathrm{Int}.
\end{align}
The first term is the Hamiltonian of the exciton-photon system, $\hat{H}_\mathrm{S}=\hbar\omega_\mathrm{ph}\hat{a}^\dagger \hat{a}+\sum_{r=\mathrm{br},\mathrm{d1},\mathrm{d2}}\varepsilon_{r}\hat{b}^\dagger_{r}\hat{b}_{r}+\hbar g(\hat{b}_{\mathrm{br}}^\dagger\hat{a}+\hat{a}^{\dagger}\hat{b}_{\mathrm{br}})$ with $\hat{a}^{\dagger}$, $\hat{b}^\dagger_{\mathrm{br}}$, and $\hat{b}^\dagger_{\mathrm{d1,d2}}$ being the bosonic creation operators for the photon, the bright exciton, and the two dark excitons (dark-1 and dark-2), respectively, where the dark-1 (dark-2) exciton has even-parity (triplet) symmetry. The bright exciton is coupled to the photon with the dipole coupling strength $g$. The second term represents the phononic bath, $\hat{H}_\mathrm{B}=\sum_{r=\mathrm{d1},\mathrm{d2}}\sum_{\bm{q}}\hbar\Omega_{r\bm{q}}\hat{p}^\dagger_{r\bm{q}}\hat{p}_{r\bm{q}}$ where $\hat{p}^\dagger_{r\bm{q}}$ reads the bosonic creation operator of a phonon with a momentum $q$. The energy of the photon, the excitons, and the phonons are represented by $\hbar\omega_\mathrm{ph}$, $\varepsilon_{r=\mathrm{d1},\mathrm{d2}}$, and $\hbar\Omega_{r\bm{q}}$, respectively. The last term indicates the phonon-mediated coupling between the bright and dark excitons, $\hat{H}_\mathrm{Int}=\sum_{r=\mathrm{d1},\mathrm{d2}}\sum_{\bm{q}}\hbar \zeta_{r\bm{q}}(\hat{b}^\dagger_{r}\hat{b}_{\mathrm{br}}+\hat{b}^\dagger_{\mathrm{br}}\hat{b}_{r})(\hat{p}^\dagger_{r\bm{q}}+\hat{p}_{r\bm{q}})$ with the effective coupling strength $\zeta$.

Simulating radiative lifetime of excitons requires non-unitary dissipations of the photon to environment or by photo detections. We simulate the non-unitary events by using quantum continuous measurement theory of photon counting with the photon dissipation rate, $\kappa\ge 0$~\cite{Srinivas:1981hi,Srinivas:1982el}. The photon counting gives the probability that $m$ photons are absorbed to environments (or disappear at the instrument) until a time $t\ge0$ after the absorption of a photon (or the creation of the bright exciton) at $t=0$: $\mathcal{P}(m;t)$.

For feasible simulations of the excitons with the photon counting, we use the moment generating function (MGF) of the probability $\mathcal{M}(\lambda;t)\equiv\sum_{m=-\infty}^{\infty}\exp[im\lambda]\mathcal{P}(m;t)$, with the conjugate variable $\lambda$ to the absorbed photon number $m$. The MGF dynamics can be explicitly determined by the following equation of motion for a generalized density matrix with $\lambda$:
\begin{align}
\partial_{t}\hat{\rho}(t;\lambda)&=-i[\hat{H},\hat{\rho}(t;\lambda)]/\hbar+\kappa\mathrm{e}^{i\lambda} \hat{a}^{}\hat{\rho}(t;\lambda)\hat{a}^\dagger\notag\\
&\quad -\kappa[\hat{a}^\dagger\hat{a}^{}\hat{\rho}(t;\lambda)+\hat{\rho}(t;\lambda)\hat{a}^\dagger\hat{a}^{}],
\end{align}
where the MGF is given by the trace of $\hat{\rho}(t;\lambda)$ as $\mathcal{M}(\lambda;t)=\Tr[\hat{\rho}(t;\lambda)]$ for $t\ge0$ with the initial boundary condition, $\hat{\rho}(t=0;\lambda)=\hat{\rho}_0$. Here, $\hat{\rho}_0$ is the initial density matrix just after the photon absorption. Applying the Franck-Condon principle to the initial state at $t=0$, the initial state is reasonably given by a separable state of the form $\hat{\rho}_{0}\propto\hat{\rho}_{0}^\mathrm{S}\otimes\exp[-\hat{H}_\mathrm{B}/k_B T]$ where $\hat{\rho}_{0}^\mathrm{S}$ is the initial density matrix of the exciton-photon system with one bright exciton and $T$ is the temperature of the phononic bath.

From now on, we assume that the coupling strength $\zeta_\mathrm{d1,d2}$ is weak and take the influence into account by the standard second-order perturbation approximation in terms of $\hat{H}_\mathrm{Int}$~\cite{Petruccione:378945}. In this approximation, the exciton dynamics is adequately described by the GQME for the generalized reduced density matrix (RDM), $\hat{\rho}^\mathrm{S}(t;\lambda)\equiv\Tr_\mathrm{B}[\hat{\rho}^{}(t;\lambda)]$ where $\Tr_\mathrm{B}$ means the partial trace of the phononic bath, and the influence of the phonons on the exciton system is represented by the spectral density $J_r(\omega)\equiv\sum_{\bm{q}}|\zeta_{r\bm{q}}|^2\delta(\omega-\Omega_{r\bm{q}})$ for $r=\mathrm{d1,d2}$. Here, we assume that $J_r(\omega)$ has the standard Ohmic form~\cite{Petruccione:378945}, $J_r(\omega)=2\gamma_r^2\omega\theta(\omega_\mathrm{cut}^{}-\omega)/\omega_\mathrm{cut}^2$ where $\gamma_r\equiv\sqrt{\sum_{\bm{q}}|\zeta_{r\bm{q}}|^2}$ and $\omega_\mathrm{cut}$ denote the effective coupling strength between the exciton and phonons and the energy cutoff of the spectral density, respectively.

Solving the GQME numerically, we calculate the generalized RDM that gives the MGF as $\mathcal{M}(\lambda;t)=\Tr_\mathrm{S}[\hat{\rho}^\mathrm{S}(t;\lambda)]$. The probability is obtained from the inverse Fourier transform of the MGF. Note that the generalized RDM $\hat{\rho}^\mathrm{S}(t;\lambda=0)$ coincides with the actual density matrix of the reduced system where we ignore the counting history, $\hat{\rho}^\mathrm{S}(t)=\hat{\rho}^\mathrm{S}(t;\lambda=0)$~\cite{Supplement,Srinivas:1981hi,Srinivas:1982el}. Hereafter, we omit the script ``S'' representing the subspace of the exciton-photon system such as $\hat{\rho}^\mathrm{S}(t)\to\hat{\rho}(t)$ for simplicity. Although the procedure is straightforward, the details of the formalism of the GQME are described in the Supplemental Material~\cite{Supplement}.

Actual time-resolved photoluminescence experiments have time-resolution limit $\varDelta t$, which is introduced into our theory phenomenologically by the Gaussian averaging: $\tilde{\mathcal{P}}(m,t)\equiv\frac{1}{\sqrt{2\pi}\varDelta t}\int_{-\infty}^{\infty}\mathcal{P}(m,\tau)\exp[-(\tau-t)^2/(\sqrt{2}\varDelta t)^2] d\tau$. From the time-averaged probability with $\varDelta t=0.1$ ps, the time-resolved photoluminescence, $L(t)$, and the quantum yield at a time $t$, $Y(t)$, are given by
\begin{align}
L(t)&\equiv \sum_{m}m\partial_{t}\tilde{\mathcal{P}}(m;t),\\
Y(t)&\equiv \sum_{m}m\tilde{\mathcal{P}}(m;t)/m_{0}.
\end{align}
where $m_{0}=1$ is the initial number of the bright exciton at $t=0$. Note that $Y(t)$ represents the maximum quantum yield of photoluminescence experiments because the other environmental-photon-absorption is also included in $Y(t)$.

For a model calculation of (6,5) SWCNTs at room temperature, we choose the value of the parameters as follows: $\hbar\omega_\mathrm{ph}=\varepsilon_\mathrm{br}=1.27$ eV, $\varepsilon_\mathrm{d1}=1.265$ eV, $\varepsilon_\mathrm{d2}=1.15$ eV, $T=300$ K, $\hbar\omega_\mathrm{cut}=0.2$ eV, $\hbar g=10.5$ meV, $\hbar\gamma_\mathrm{d1}=0.875$ meV, and $\hbar\gamma_\mathrm{d2}=0.25$ meV. Effects of the environment are described by the single parameter, namely, the decay rate of photon, $\kappa$. Since $\kappa$ strongly depends on the ambient solvents, matrices, and/or substrates, we vary $\kappa$ over several orders of magnitude with keeping $\kappa$ as a small parameter. The energy levels of bright and dark excitons are estimated from the photoluminescence experiments~\cite{Shaver:2008mo,Berciaud:2008ld,Matsunaga:2008ef,Matsunaga:2010ex,Nagatsu:2010bo}. The cutoff frequency of the spectral density $\hbar\omega_\mathrm{cut}$ is determined from the density of states of phonons in SWCNTs~\cite{Dresselhaus:2000ge}. The coupling strength $\gamma_{\mathrm{d1,d2}}$ is determined from the numerical fitting that reproduces the experiment~\cite{Duque:2009gs} (see Fig.~\ref{fig3}(c)). The dipole coupling strength $g$ is taken to be $g\gg\gamma_{\mathrm{d1,d2}}$. Note that one confirms that one order of magnitude difference in $g$ does not affect our main results.

\begin{figure}[thbp]
\centering
\includegraphics[keepaspectratio,width=8.6cm]{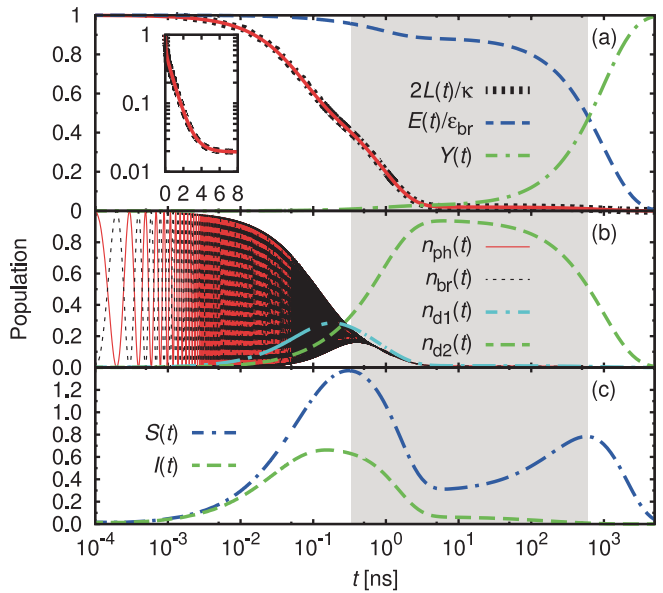}
\caption{(Color online) Transient exciton dynamics in a model of (6,5) SWCNT. (a) Time-resolved photoluminescence $L(t)$ and quantum yield $Y(t)$. The parameter of radiative dissipation to environment is fixed at $\kappa^{-1}=10$ ns. The solid red line indicates the tri-exponential fitting of the numerical result, $L_{\mathrm{fit}}(t)$. The time-dependent energy $E(t)$ is also plotted in the same figure. (b) Population dynamics of the exciton-photon system with $n_{\mathrm{ph}}(t)\equiv\Tr[\hat{\rho}(t)\hat{a}^\dagger\hat{a}^{}]$ and $n_{r=\mathrm{br},\mathrm{d1},\mathrm{d2}}(t)\equiv\Tr[\hat{\rho}(t)\hat{b}^\dagger_{\mathrm{r}}\hat{b}_{\mathrm{r}}^{}]$. (c) Time evolution of von Neumann entropy $S$ and the quantum mutual information $I$. The shadow region indicates the interval between the two peaks of the von Neumann entropy.}
\label{fig2}
\end{figure}

First, we show the time-resolved photoluminescence $L(t)$ for $\kappa^{-1}=10$ ns in Fig.~\ref{fig2}(a), which may be a typical result for SWCNTs in aqueous solutions. The simulated $L(t)$ is accurately fitted by a tri-exponential function:
\begin{align}
L_{\mathrm{fit}}(t)=l_{1}\exp(-t/\tau_1)+l_{2}\exp(-t/\tau_2)+l_{3}\exp(-t/\tau_3)
\end{align}
with $\tau_1<\tau_2<\tau_3$. We note that similar multi-exponential curves are also observed in the experiments~\cite{Kono:2004uf,Berciaud:2008ld,Duque:2009gs,Miyauchi:2009rl}. The decay constants obtained in our calculation differ by several orders of magnitude: The fast decay is characterized by $\tau_1$ $=$ 65 ps while the intermediate and slow decay constants, $\tau_2$ and $\tau_3$, are found at 890 ps and 1 $\mu$s, respectively.

While the luminescence rapidly decreases within 1 ns as shown in Fig.~\ref{fig2}(a), the energy of the exciton-photon system $E(t)\equiv\Tr[\hat{\rho}(t)\hat{H}_\mathrm{S}]$ remains over 80\% of $\varepsilon_\mathrm{br}$ even at 100 ns with the plateau from 5ns to 50 ns where the quantum yield holds lower than 10\%. The significant difference in the decays of energy and luminescence qualitatively explains both the seemingly contradictory experiments of the time-resolved photoluminescence~\cite{Berciaud:2008ld,Duque:2009gs,Stich:2014kr} and the pump-probe transient absorption spectroscopy~\cite{Park:2011jf,Stich:2014kr}: Some time-resolved photoluminescence spectroscopies show the rapid luminescence decay with tens of ps and a low quantum yield~\cite{Berciaud:2008ld,Duque:2009gs,Stich:2014kr} whereas a dark exciton survives over 10 $\mu$s in the pump-probe transient absorption spectroscopy~\cite{Park:2011jf,Stich:2014kr}.

Exciton lifetime is a crucial factor characterizing the dynamics. Here, we define the energy lifetime of excitons $\tau_\mathrm{LT}$ as the energy decay time to $1/e$ of the initial value for consistency with the single-exponential decay model:
\begin{align}
E(\tau_\mathrm{LT})/E(0)\equiv e^{-1}.
\end{align}
From the definition, we obtain $\tau_\mathrm{LT}=880$ ns, which is longer than forty times of that in the system with no dark exciton given by the inverse of the rate of the photon absorption to the environment $\tau_\mathrm{LT}^\mathrm{no dark}=2\kappa^{-1}=20$ ns as detailed later. Here, the factor ``two'' in $2\kappa^{-1}$ results from the halved residence time of the photon state by the Rabi oscillation between the photon and the bright exciton~\cite{Norris:1994tr}.

The enhanced energy lifetime of the exciton-photon system is attributed to the fast irreversible relaxation pathway from the bright exciton to the dark excitons accompanied by quantum-to-classical crossover. We see the relaxation in the population dynamics in Fig.~\ref{fig2}(b). The initial dynamics has a quantum nature where the populations of the photon and the bright exciton are oscillating with the anti-phase with the frequency of $g/2$, which clearly indicates the Rabi oscillation. Then, the population is gradually transferred to the dark excitons after 10 ps with a reduction of the oscillation due to decoherence effects by phonons. In particular, the dark-2 exciton is stabilized from a few ns to 100 ns with a high population via the quantum-to-classical crossover.

The irreversible crossover is confirmed by using the time-dependence of the von Neumann entropy and the quantum mutual information as shown in Fig.~\ref{fig2}(c). The von Neumann entropy, $S(t)\equiv -\Tr[\hat{\rho}(t)\ln \hat{\rho}(t)]$, initially remains small, which means that the initial dynamics is dominated by the time-evolution of a quantum pure state. Then, the von Neumann entropy increases due to the decoherence by phonons, and shows two peaks approximately at 300 ps and 600 ns with the strong mixing of states. These mixings, however, have different origins, {\it i.e.} the correlation between particles including quantum entanglement at $\sim 300$ ps and the classical stochastic mixing of the state at $\sim 600$ ns.

To distinguish these two distinct origins of the mixing, we introduce quantum mutual information: The quantum mutual information provides a measure of correlation between subsystems of quantum states. Here, we calculate the quantum mutual information between the subsystem X consisting of the photon and the bright exciton and the subsystem Y of the dark excitons, $I(t)\equiv S(\hat{\rho}(t)\|\hat{\rho}_{\mathrm{X}}(t)\otimes\hat{\rho}_{\mathrm{Y}}(t))$. Here $S({\hat{\rho}\|\hat{\sigma}})\equiv\Tr[\hat{\rho}(\ln(\hat{\rho})-\ln(\hat{\sigma}))]$ is the quantum relative entropy that gives a measure of difference between two quantum states, and $\hat{\rho}_{\mathrm{X}}(t)$ and $\hat{\rho}_{\mathrm{Y}}(t)$ are the RDMs defined by the partial trace of the subsystem of Y and X, respectively: $\hat{\rho}_{\mathrm{X}}(t)\equiv\Tr_{\mathrm{Y}}[\hat{\rho}(t)]$ and $\hat{\rho}_{\mathrm{Y}}(t)\equiv\Tr_{\mathrm{X}}[\hat{\rho}(t)]$.

While the correlation between the subsystems is enhanced approximately at the first peak of $S(t)$, $t\simeq$300 ps, the quantum mutual information gives no peak at the second peak of $S(t)$, which implies that the system becomes a nearly separable state with essentially no quantum entanglement after the first peak as shown in Fig.\ref{fig2}(c). Hence, the crossover occurs near the first peak , which leads to the superselection of dark exciton states imposed by the decoherence~\cite{Zurek:2003de}. Ultimately, the fast relaxation from the bright exciton to the dark excitons and the rapid quantum-to-classical crossover do not contradict each other in the time domain.

\begin{figure}[thbp]
\centering
\includegraphics[keepaspectratio,width=8.6cm]{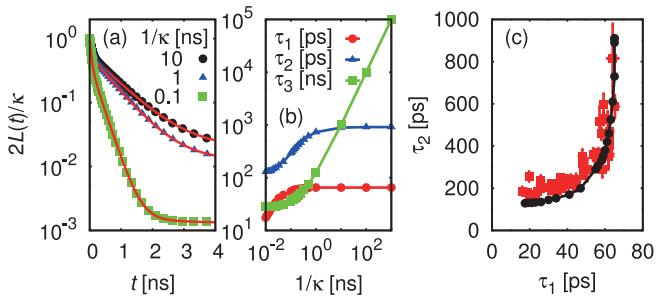}
\caption{(Color online) Environmental effects on photoluminescence. (a) Time-resolved photoluminescence with several choices of $\kappa^{-1}$. The solid red lines indicate the tri-exponential fitting functions on the numerical results. (b) $\kappa$-dependence of decay constants in tri-exponential fitting function. (c) Correlation between $\tau_1$ and $\tau_2$ for several choices of $\kappa^{-1}$ , which ranges from 10 ps to 1 $\mu$s (black filled circle). The filled square (red) symbols indicate the experimentally measured decay constants for (6,5) SWCNT reproduced from Ref.~\cite{Duque:2009gs}.}
\label{fig3}
\end{figure}

Next, we examine the environmental effects on the time-resolved photoluminescence by monitoring $\kappa$ as shown in Fig.~\ref{fig3}(a). The photoluminescence decays are well fit by a tri-exponential function irrespective of $\kappa$ though the decay constants strongly depend on $\kappa$. The two decay constants $\tau_1$ and $\tau_2$ for the fast and intermediate decays, respectively, are saturated with an increase in $\kappa^{-1}$ as shown in Fig.~\ref{fig3}(b). Besides, the largest decay constant $\tau_3$ shows no saturation and proportional to $100\kappa^{-1}$ with increasing $\kappa^{-1}$. These decay constants share the common origin: The positive correlation between $\tau_1$ and $\tau_2$ is clearly seen in Fig.~\ref{fig3}(c). The convex curve is consistent with the results of the experiment~\cite{Duque:2009gs}.

\begin{figure}[thbp]
\centering
\includegraphics[keepaspectratio,width=8.6cm]{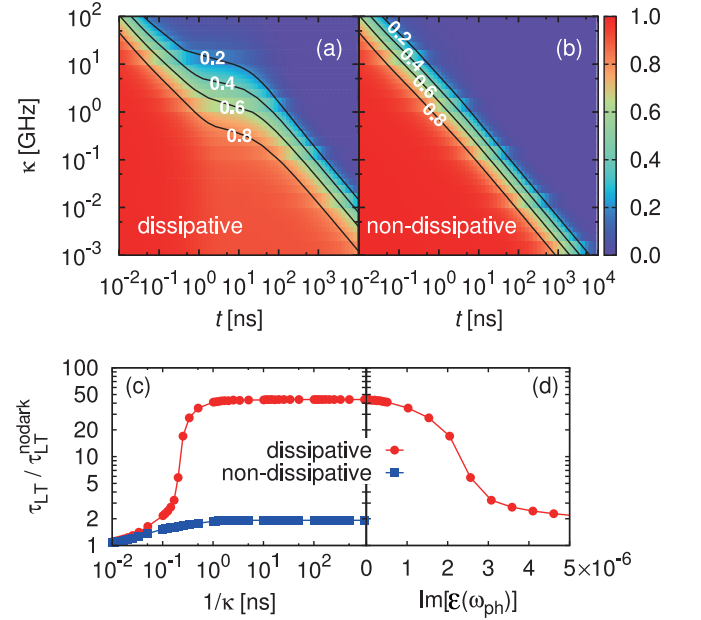}
\caption{(Color online) Environmental effects on energy relaxation. (a,b) Normalized energy $E/\varepsilon_\mathrm{br}$ as a function of time and $\kappa$ for the (6,5) SWCNT model (dissipative system), and for the hypothetical model with the same parameters as those in the dissipative system except for the energy levels, $\hbar\omega_\mathrm{ph}=\varepsilon_{\mathrm{br}}=\varepsilon_{\mathrm{d1}}=\varepsilon_{\mathrm{d2}}$ (non-dissipative system). Solid lines indicate the contours of the normalized energy. (c) Energy lifetime $\tau_\mathrm{LT}$ as a function of $\kappa$ in the dissipative and non-dissipative systems. The lifetime is normalized by $\tau_\mathrm{TL}^\mathrm{nodark}=2\kappa^{-1}$ that is the lifetime in the system with no dark exciton. (d) Normalized lifetime in the dissipative system as a function of the imaginary part of the dielectric function at $\omega=\omega_\mathrm{ph}$ of the ambient dielectric medium, $\mathrm{Im}[\varepsilon(\omega_\mathrm{ph})]=\kappa/\omega_\mathrm{ph}$.}
\label{fig4}
\end{figure}

Finally, we emphasize the importance of the energy dissipation for the enhancement of the exciton lifetime. Figures~\ref{fig4}(a) and (b) show the system energy $E(t)$ as a function of time and $\kappa$ for the systems where (a) the dark excitons have the lower energy compared to the bright one (dissipative system) or (b) all the excitons have the same energy level (non-dissipative system). The dissipative system obviously survives longer than the non-dissipative system for small $\kappa$ where the irreversible relaxation becomes dominant at the initial dynamics for $t\lesssim$100 ps.

The difference is quantified in the energy lifetime $\tau_\mathrm{LT}$ as shown in Fig.~\ref{fig4}(c). Although the lifetime of the non-dissipative system is enhanced compared to that in the system with no dark exciton (no-dark system) $\tau_\mathrm{LT}^\mathrm{no dark}=2\kappa^{-1}$, the enhancement factor is not large even for large $\kappa^{-1}$: The factor is about two. While, the lifetime in the dissipative system is about 45 times as long as $\tau_\mathrm{LT}^\mathrm{no dark}$ for large $\kappa^{-1}$. The large enhancement of the lifetime suddenly occurs within a small range of $\kappa$. When all the energy loss to the environment can be represented by the dielectric loss of the ambient homogeneous linear dielectric medium, $\kappa$ is related with the background dielectric function $\varepsilon(\omega)$ with $\omega$ being a frequency by using the Poyinting's theorem in the classical electrodynamics~\cite{Jackson:490457}: $\kappa=\omega_\mathrm{ph}\mathrm{Im}[\varepsilon(\omega_\mathrm{ph})]$. In the case, the sudden enhancement of lifetime is estimated to occur at $\mathrm{Im}[\varepsilon(\omega_\mathrm{ph})]\simeq 2 \times 10^{-6}$ as shown in the inset of Fig.~\ref{fig4}(c) , which is comparable to that of water in room temperature, $\mathrm{Im}[\varepsilon(\omega_\mathrm{ph})]\simeq1\times10^{-5}$~\cite{Palmer:1974ta}. Therefore, the exciton lifetime of the SWCNT placed in the medium with smaller $\mathrm{Im}[\varepsilon(\omega_\mathrm{ph})]$, \textit{e.g.} air, can be strongly enhanced.

In summary, we have investigated the effect of dissipation and decoherence on the long-term exciton dynamics in a model of the semiconducting (6,5) carbon nanotube. Counterintuitively, the dissipation and decoherence enhance the exciton lifetime in the relaxation pathway through the dark excitons, which paradoxically appears in experiments as a rapid decay of photoluminescence and a slow growth of quantum yield. In the search for efficient solar cells, such bad radiators are at first glance unpromising. However, we have shown this is not necessarily true. Our study confirms that semiconducting carbon nanotubes become potentially efficient photovoltaic materials as suggested in the experiments~\cite{Arnold:2013rd}, and provides a further guideline and insight into high potentials of other unexplored materials.

\begin{acknowledgements}
We acknowledge the financial supports by a Grant-in-Aid for Scientific Research (No. 22104010 and No. 22340090) from MEXT, Japan, and a Grant-in-Aid for Scientific Research on Innovative Areas `Materials Design through Computics: Complex Correlation and Non-Equilibrium Dynamics'. This work was also supported by MEXT HPCI Strategic Programs for Innovative Research (SPIRE) (under the grant number hp130007 and hp140215) and Computational Materials Science Initiative (CMSI).
\end{acknowledgements}

\clearpage
\renewcommand{\thefigure}{S\arabic{figure}}
\renewcommand{\theequation}{S\arabic{equation}}
\setcounter{figure}{0}
\setcounter{equation}{0}

\section{Supplemental Material}
We describe the details of our method to calculate the density matrix $\hat{\rho}(t)$ and the probability $\mathcal{P}(m;t)$ referred to in the main article. Note that we use the notation of the doubled Hilbert space, {\it i.e.} the Liouville space, where a matrix $\hat{\sigma}$ in the normal Hilbert space becomes a vector $\kket{\hat{\sigma}}$ with the Hilbert-Schmidt inner product, $\bbrakket{\hat{\rho}}{\hat{\sigma}}\equiv\Tr[\hat{\rho}^\dagger\hat{\sigma}]$~\cite{Arimitsu:1987df:sm,Esposito:2009fd:sm}. A linear matrix operation becomes a linear map on the Liouville space, {\it i.e.} a superoperator. This description is useful for both the formulation and the numerical calculation in open quantum systems.

Let us first consider the closed exciton system [Eq. (1) in the main article] with no interaction with the ambient environment. In this case, the dynamics of the initial density matrix $\kket{\hat{\rho}_0}$ is fully determined by the system Hamiltonian $\hat{H}$. The time-evolution during the time interval $[0,t]$ is given by the unitary superoperator,
\begin{align}
\breve{U}(t)\equiv\exp[-i[\breve{\mathfrak{L}}(\hat{H})-\breve{\mathfrak{R}}(\hat{H})]t/\hbar],
\end{align}
where $\breve{\mathfrak{L}}(\cdot)$ ($\breve{\mathfrak{R}}(\cdot)$) is the map from any operator $\hat{A}$ in a Hilbert space to the superoperator $\breve{\mathfrak{L}}(\hat{A})$ ($\breve{\mathfrak{R}}(\hat{A})$) in the Liouville space that makes the operator act to a matrix from left (right): $\breve{\mathfrak{L}}(\hat{A})\kket{\hat{\sigma}}=\kket{\hat{A}\hat{\sigma}}$ and $\breve{\mathfrak{R}}(\hat{A})\kket{\hat{\sigma}}=\kket{\hat{\sigma}\hat{A}}$. The density matrix after the time-evolution $\kket{\hat{\rho}(t)}=\breve{U}(t)\kket{\hat{\rho}_0}$ is consistent with the trivial results
\begin{align}
\kket{\hat{\rho}(t)}&=\breve{\mathfrak{L}}(\exp[-i\hat{H}t])\breve{\mathfrak{R}}(\exp[i\hat{H}t])\kket{\hat{\rho}_0}\notag\\
&=\kket{\exp[-i\hat{H}t]\hat{\rho}_0\exp[i\hat{H}t]}.
\label{eq:unitaryevolution}
\end{align}
Note that in the first line in Eq.~\eqref{eq:unitaryevolution}, we use the commutative property between the superoperators mapped by $\breve{\mathfrak{L}}(\cdot)$ and $\breve{\mathfrak{R}}(\cdot)$, $\breve{\mathfrak{L}}(\hat{A})\breve{\mathfrak{R}}(\hat{B})=\breve{\mathfrak{R}}(\hat{B})\breve{\mathfrak{L}}(\hat{A})$, and their homomorphic and antihomomorphic properties: $\breve{\mathfrak{L}}(\hat{A}\hat{B})=\breve{\mathfrak{L}}(\hat{A})\breve{\mathfrak{L}}(\hat{B})$ and $\breve{\mathfrak{L}}(\hat{A}+\hat{B})=\breve{\mathfrak{L}}(\hat{A})+\breve{\mathfrak{L}}(\hat{B})$, and $\breve{\mathfrak{R}}(\hat{A}\hat{B})=\breve{\mathfrak{R}}(\hat{B})\breve{\mathfrak{R}}(\hat{A})$ and $\breve{\mathfrak{R}}(\hat{A}+\hat{B})=\breve{\mathfrak{R}}(\hat{A})+\breve{\mathfrak{R}}(\hat{B})$.

From now on, we shall consider the time-evolution in the open exciton system with photon dissipation to the environment based on the continuous measurement theory of photon counting~\cite{Srinivas:1981hi:sm,Srinivas:1982el:sm}. Assuming that the environment does not absorb more than one photon during any infinitesimal time interval, we consider here only two fundamental processes: no-absorption and one-photon absorption. The one-photon absorption process during any infinitesimal time interval $\delta t$ is represented by the superoperator $\breve{J}\delta t\equiv\kappa \delta t \breve{\mathfrak{L}}(\hat{a})\breve{\mathfrak{R}}(\hat{a}^\dagger)$ with $\kappa\ge0$ being the absorption rate in the presence of a free photon, which converts an $n$-photon state to an $(n-1)$-photon state~\cite{Srinivas:1981hi:sm,Srinivas:1982el:sm}. Note that $\breve{J}\delta t$ is an adequate quantum dynamical map because it is linear and completely positive (CP)~\cite{Sudarshan:1961hs:sm,Davies:1970cf:sm,Kraus:1971dv:sm}. Since the normalized pre-absorption state $\hat{\rho}_{-}$ is converted to the non-normalized post-absorption state $\hat{\rho}_+$ by the immediate absorption, $\kket{\hat{\rho}_+}=\breve{J}\delta t\kket{\hat{\rho}_{-}}$, the associated absorption probability is given by $\Tr[\hat{\rho}_+]=\bbra{\hat{I}}\breve{J}\delta t\kket{\hat{\rho}_{-}}=\kappa\delta t\Tr[\hat{a}^\dagger\hat{a}\hat{\rho}_{-}]$.

Next, we consider the no-absorption process. When the superoperator $\breve{W}_{0}(t)$ represents the time-evolution under the condition that no photon is absorbed in the interval $[0,t]$, the map $\breve{W}_{0}(\cdot)$ should be continuous and homomorphic because the no-absorption process during $[0,t]$ must be divided into a sequential process of several no-absorption processes:
\begin{align}
\breve{W}_{0}(0)&=\breve{I},\label{eq:boundary_W0}\\
\breve{W}_{0}(t_1+t_2)&=\breve{W}_{0}(t_1)\breve{W}_{0}(t_2),\label{eq:homomorphic_W0}
\end{align}
for all $t_1, t_2\ge0$, where $\breve{I}$ is the identity superoperator. These properties are the same as those of the unitary time-evolution $\breve{U}(t)$. The no-absorption process is, however, different from $\breve{U}(t)$ due to the presence of the environment. The canonical form is given by~\cite{Srinivas:1981hi:sm,Srinivas:1982el:sm}
\begin{align}
\breve{W}_{0}(t)&=\exp[-i[\breve{\mathfrak{L}}(\hat{\mathcal{H}})-\breve{\mathfrak{R}}(\hat{\mathcal{H}}^\dagger)]t/\hbar],\\
\mathcal{H}&\equiv \hat{H}-i\hbar\kappa\hat{a}^\dagger\hat{a}/2.
\end{align}
The anti-Hermitian part of $\mathcal{H}$ ensures a decrease in the associated no-absorption probability in the infinitesimal time interval $\delta t$, $\bbra{\hat{I}}\breve{W}_{0}(\delta t)\kket{\hat{\rho}_-}=1-\kappa \delta t \Tr[\hat{a}^\dagger\hat{a}\hat{\rho}_-]$, which compensates for an increase in the photon-absorption probability $\bbra{\hat{I}}\breve{J}\delta t\kket{\hat{\rho}_{-}}=\kappa\delta t\Tr[\hat{a}^\dagger\hat{a}\hat{\rho}_{-}]$.
It is noteworthy that $\breve{W}_{0}(t)$ is the CP superoperator that satisfies Eq.~\eqref{eq:boundary_W0} and Eq.~\eqref{eq:homomorphic_W0}, and approaches $\breve{U}(t)$ in the limit of $\kappa\to0$.

The actual time-evolution superoperator $\breve{W}(t)$ is defined such that it represents a time-evolution of a density matrix during the time interval $[0,t]$ with no information about the absorption. Accordingly, it is given by the sum of all possible combinations of the two fundamental processes:
\begin{align}
\breve{W}(t)&=\breve{W}_0(t)+\sum_{m=1}^{\infty}\breve{W}_{m}(t),\\
\breve{W}_{m\ge1}(t)&\equiv\int_{0}^{t}dt_{m}\cdots\int_{0}^{t_2}dt_{1}\notag\\
&\times\breve{W}_0(t-t_m)\breve{J}\cdots \breve{W}_0(t_2-t_1)\breve{J}\breve{W}_0(t_1),
\end{align}
where $\breve{W}_{m}(t)$ represents the $m$-photon absorption process during the time interval $[0,t]$. The actual time-evolution is also described by the recursive relation,
\begin{align}
&\breve{W}(t)=\breve{W}_0(t)+\int_{0}^{t}d\tau\breve{W}(t-\tau)\breve{J}\breve{W}_0(\tau),
\end{align}
which clearly indicates that the actual time-evolution consists of the immediate one-photon absorption process and the no-absorption process. In addition, $\breve{W}(t)$ satisfies the following equation-of-motion (EOM):
\begin{align}
\partial_{t}\breve{W}(t)&=[\breve{\mathcal{L}}_\mathrm{S}+\breve{\mathcal{L}}_\mathrm{B}+\breve{\mathcal{L}}_\mathrm{Int}]\breve{W}(t).
\label{eq:timeevolution_W}
\end{align}
where the Liouvillians are defined by $\breve{\mathcal{L}}_\mathrm{S}\equiv -i[\breve{\mathfrak{L}}(\hat{H}_\mathrm{S})-\breve{\mathfrak{R}}(\hat{H}_\mathrm{S})]/\hbar+\kappa\breve{\mathfrak{L}}(\hat{a})\breve{\mathfrak{R}}(\hat{a}^\dagger)-\kappa[\breve{\mathfrak{L}}(\hat{a}^{\dagger}\hat{a})+\breve{\mathfrak{R}}(\hat{a}^{\dagger}\hat{a})]/2$, $\breve{\mathcal{L}}_\mathrm{B}\equiv -i[\breve{\mathfrak{L}}(\hat{H}_\mathrm{B})-\breve{\mathfrak{R}}(\hat{H}_\mathrm{B})]/\hbar$, and $\breve{\mathcal{L}}_\mathrm{Int}\equiv-i[\breve{\mathfrak{L}}(\hat{H}_\mathrm{Int})-\breve{\mathfrak{R}}(\hat{H}_\mathrm{Int})]/\hbar$, respectively. Here $\hat{H}_{\mathrm{S,B,Int}}$ is defined in Eq. (1) of the main article. Accordingly, we can obtain the density matrix after the actual time-evolution $\kket{\hat{\rho}(t)}=\breve{W}(t)\kket{\hat{\rho}_0}$ by solving the EOM. Note that the EOM has a Lindblad form~\cite{Petruccione:378945:sm} and $\kket{\hat{\rho}(t)}$ is normalized as $\Tr[\hat{\rho}(t)]=\bbrakket{\hat{I}}{\hat{\rho}(t)}=\bbra{\hat{I}}\breve{W}(t)\kket{\hat{\rho}_0}=\Tr[\hat{\rho}_0]$. It is noted that if there is a photon decoupled from the excitons at $t=0$, the population decays exponentially with the lifetime of $\kappa^{-1}$, which is calculated analytically~\cite{Srinivas:1981hi:sm}.

Let us turn to the calculation of probability $\mathcal{P}(m;t)$ that $m$-photons are absorbed during the time interval $[0,t]$. From the above discussion, we have already obtained the superoperator that represents the $m$-photon absorption process during $[0,t]$ $\breve{W}_{m}(t)$:
\begin{align}
\mathcal{P}(m;t)=\bbra{\hat{I}}\breve{W}_{m}(t)\kket{\hat{\rho}_0}.
\end{align}
Although it is difficult to calculate the probability analytically due to the phonon interaction, we are able to numerically calculate it with an approximation based on the moment generating function (MGF) of the probability,
\begin{align}
&\mathcal{M}(\lambda;t)\equiv\sum_{m=0}^\infty\mathrm{e}^{i\lambda m}\mathcal{P}(m;t)=\bbrakket{I}{\hat{\rho}(t;\lambda)},
\end{align}
where we define the generalized density matrix by $\kket{\hat{\rho}(t;\lambda)}\equiv\breve{W}(t;\lambda)\kket{\hat{\rho}_0}$ with $\breve{W}(t;\lambda)\equiv\breve{W}(t)|_{\breve{J}\to\breve{J}\exp(i\lambda)}$ being the time-evolution superoperator. The EOM for $\kket{\hat{\rho}(t;\lambda)}$ is therefore given by the following equation,
\begin{align}
\partial_{t}\kket{\hat{\rho}(t;\lambda)}&=[\breve{\mathcal{L}}_\mathrm{S}(\lambda)+\breve{\mathcal{L}}_\mathrm{B}+\breve{\mathcal{L}}_\mathrm{Int}]\kket{\hat{\rho}(t;\lambda)},
\label{eq:timeevolution_GDM}
\end{align}
with $\breve{\mathcal{L}}_\mathrm{S}(\lambda)\equiv\breve{\mathcal{L}}_\mathrm{S}|_{\breve{J}\to\breve{J}\exp(i\lambda)}$. The boundary condition is given by the actual density matrix at $t=0$: $\kket{\hat{\rho}(0;\lambda)}=\kket{\hat{\rho}_0}$. Here, the initial density matrix $\kket{\hat{\rho}_0}$ is assumed to be the tensor product state of a bright exciton state and the phonon bath in the thermal equilibrium state, which reasonably describes the state just after the exciton generation in accordance with the Franck-Condon principle: $\kket{\hat{\rho}_0}=\kket{\hat{\rho}^\mathrm{S}_{0}}_\mathrm{S}\otimes \kket{\hat{\rho}^\mathrm{B}_\mathrm{eq}}_\mathrm{B}$ where $\hat{\rho}^\mathrm{S}_{0}\equiv \hat{b}^{\dagger}_{\mathrm{br}}\ket{0}\bra{0}\hat{b}^{}_{\mathrm{br}}$ and $\hat{\rho}^\mathrm{B}_\mathrm{eq}\equiv \exp[-\hat{H}_\mathrm{B}/k_{B}T]/\Tr_\mathrm{B}[\exp[-\hat{H}_\mathrm{B}/k_{B}T]]$. Here, $\Tr_\mathrm{B}$ is the partial trace with respect to the subspace of the phonon bath, $k_{B}$ is the Boltzmann constant, and $T$ is the temperature of the bath.

Equation~\eqref{eq:timeevolution_GDM} is equivalent to Eq.~(2) in the main article, which is written in the ordinary Hilbert space. Because the actual time-evolution of the density matrix is given by $\kket{\hat{\rho}(t)}=\kket{\hat{\rho}(t;0)}$ [compare Eq.~\eqref{eq:timeevolution_W} and Eq.~\eqref{eq:timeevolution_GDM}], we hereafter focus on the calculation of the EOM~\eqref{eq:timeevolution_GDM}. The probability $\mathcal{P}(m;t)$ is given by the inverse Fourier transformation of the MGF.

In the main article, we assume that the coupling between the phonons and the excitons is weak. Hence, we include the phonon effects on the exciton-photon system within the second order perturbation in terms of the interaction Liouvillian $\breve{\mathcal{L}}_\mathrm{Int}$. The phonon bath is also assumed to remain in the thermal equilibrium at all time.

Because the time-evolution equation \eqref{eq:timeevolution_GDM} has a form very similar to that for the standard density matrix, we make use of an established method developed in the quantum master equation (QME). Then, we derive a closed EOM for the generalized reduced density matrix in the exciton-photon subspace $\kket{\hat{\rho}_{}^{\rm S}(t;\lambda)}_{\rm S}$ defined by
\begin{align}
\kket{\hat{\rho}_{}^{\rm S}(t;\lambda)}_{\rm S}&\equiv{}_{\rm B}^{}\bbrakket{\hat{I}_{\rm B}}{\hat{\rho}_{}(t;\lambda)},
\end{align}
where ${}_{\rm B}^{}\bbra{\hat{I}_{\rm B}}$ means the partial trace of the phonon bath subspace ${}_{\rm B}^{}\bbrakket{\hat{I}_{\rm B}}{\hat{\rho}}\equiv\kket{\Tr_{\rm B}[\hat{I}_{\rm B}^\dagger\hat{\rho}]}_{\rm S}=\kket{\Tr_{\rm B}[\hat{\rho}]}_{\rm S}$.
In accordance with the standard second-order time-convolutionless projection operator method~\cite{Petruccione:378945:sm}, we start with the density matrix in the interaction picture,
\begin{align}
\kket{\hat{\rho}_{\mathrm{I}}(t;\lambda)}\equiv \exp[-(\breve{\mathcal{L}}_\mathrm{S}(\lambda)+\breve{\mathcal{L}}_\mathrm{B})t]\kket{\hat{\rho}(t;\lambda)}
\end{align}
Because $\kket{\hat{\rho}_{\mathrm{I}}(t;\lambda)}$ satisfies the following EOM,
\begin{align}
\partial_{t}\kket{\hat{\rho}_{\mathrm{I}}(t;\lambda)}=\breve{\mathcal{L}}^\mathrm{I}_\mathrm{Int}(t;\lambda)\kket{\hat{\rho}_\mathrm{I}(t;\lambda)},
\end{align}
with $\breve{\mathcal{L}}^\textrm{I}_\textrm{Int}(t;\lambda)\equiv\exp[-(\breve{\mathcal{L}}_\mathrm{S}(\lambda)+\breve{\mathcal{L}}_\mathrm{B})t]\breve{\mathcal{L}}_\textrm{Int}(t)\exp[(\breve{\mathcal{L}}_\mathrm{S}(\lambda)+\breve{\mathcal{L}}_\mathrm{B})t]$, the formal solution is given by
\begin{align}
\kket{\hat{\rho}_{\textrm{I}}(t;\lambda)}&=\breve{T}\exp[\int_{0}^{t}\breve{\mathcal{L}}^\textrm{I}_\textrm{Int}(\tau;\lambda)d\tau]\kket{\hat{\rho}_0}.
\end{align}
Introducing the explicit coupling constant $\alpha$ to the Liouvillian, $\alpha\breve{\mathcal{L}}^\textrm{I}_\textrm{Int}(\tau;\lambda)$, we expand the above formal solution within the second order~\cite{Petruccione:378945:sm}. Then, we obtain the following EOM for the generalized reduced density matrix in the interaction picture $\kket{\hat{\rho}_{\textrm{I}}^{\rm S}(t;\lambda)}_{\rm S}\equiv{}_{\rm B}^{}\bbrakket{\hat{I}_{\rm B}}{\hat{\rho}_{\textrm{I}}(t;\lambda)}$,
\begin{align}
\partial_{t}\kket{\hat{\rho}_{\textrm{I}}^{\rm S}(t;\lambda)}_{\rm S}=\exp[-\breve{\mathcal{L}}_\mathrm{S}(\lambda)t]\breve{\Upsilon}(t;\lambda)\exp[\breve{\mathcal{L}}_\mathrm{S}(\lambda)t]\kket{\hat{\rho}_{\textrm{I}}^{\rm S}(t;\lambda)}_{\rm S},
\end{align}
where we define
\begin{align}
&\breve{\Upsilon}(t;\lambda)\equiv-\int_{0}^{t}d\tau\int_{0}^{\infty} d\omega\sum_{r=\mathrm{d1},\mathrm{d2}} J_{r}(\omega)\breve{B}_{r}^{-}\notag\\
&\times[\mathrm{coth}(\frac{\hbar\omega}{2k_{B}T})\cos(\omega \tau)\breve{B}_{r}^{-}(\tau;\lambda)-i\sin(\omega \tau)\breve{B}_{r}^{+}(\tau;\lambda)].
\end{align}
Here, $\breve{B}_{r}^{\pm}\equiv\breve{\mathfrak{L}}(\hat{b}^\dagger_{r}\hat{b}_{\mathrm{br}}+\hat{b}^\dagger_{\mathrm{br}}\hat{b}_{r})\pm\breve{\mathfrak{R}}(\hat{b}^\dagger_{r}\hat{b}_{\mathrm{br}}+\hat{b}^\dagger_{\mathrm{br}}\hat{b}_{r})$ and $\breve{B}_{r}^{\pm}(t;\lambda)=\exp[\breve{\mathcal{L}}_{\mathrm{S}}(\lambda)t]\breve{B}_{r}^{\pm}\exp[-\breve{\mathcal{L}}_{\mathrm{S}}(\lambda)t]$. The influence of phonons on the excitons is represented by the spectral density $J_{r}(\omega)\equiv\sum_{\bm{q}}|\zeta_{r\bm{q}}|^2\delta(\omega-\Omega_{r\bm{q}})$ for $r=\mathrm{d1,d2}$, which is here assumed to have an Ohmic form; $J_{r}(\omega)=2\gamma_{r}^2\omega\theta(\omega)\theta(\omega_\mathrm{cut}-\omega)/\omega_\mathrm{cut}^2$ where we introduce a cutoff frequency $\omega_\mathrm{cut}$ and the effective coupling strength $\gamma_{r}$.

Converting the equation in the interaction picture to that in the Schr\"odinger picture, we obtain the generalized QME for numerical calculation:
\begin{align}
\partial_{t}\kket{\hat{\rho}^{\rm S}(t;\lambda)}_{\rm S}=[\breve{\mathcal{L}}_{\textrm{S}}(\lambda)+\breve{\Upsilon}(t;\lambda)]\kket{\hat{\rho}^{\rm S}(t;\lambda)}_{\rm S}.
\label{eq:timeevolution_RGDM}
\end{align}
The initial condition is given by $\kket{\hat{\rho}^{\rm S}(0;\lambda)}_{\rm S}=\kket{\hat{\rho}^\mathrm{S}_{0}}_\mathrm{S}$.
With using $\hat{\rho}^\mathrm{S}(t;\lambda)$, the MGF and the actual reduced density matrix of the exciton-photon subsystem $\hat{\rho}^\mathrm{S}(t)$ are calculated as
\begin{align}
\hat{\rho}^\mathrm{S}(t)&=\hat{\rho}^\mathrm{S}(t;0),\\
\mathcal{M}(\lambda;t)&=\bbrakket{\hat{I}}{\hat{\rho}(t;\lambda)}={}_\mathrm{S}\bbrakket{\hat{I}_\mathrm{S}}{\hat{\rho}^{\mathrm{S}}(t;\lambda)}_{\mathrm{S}}.
\end{align}

We solve the generalized QME~\eqref{eq:timeevolution_RGDM} numerically. In the Liouville space where a density matrix is vectorized and superoperators can be represented by matrices, the QME~\eqref{eq:timeevolution_RGDM} becomes just a system of linear differential equations. Hence, we use ordinary methods: Fourth-order Runge-Kutta (RK4) method and eigen-decomposition (ED). At the initial time, $\breve{\Upsilon}(t;\lambda)$ strongly depends on the time $t$. Therefore, we solve the equation by the RK4 method until $\breve{\Upsilon}(t;\lambda)$ becomes time-independent. We take the unit time step typically as $0.005$ ps. After sufficiently long time, $\breve{\Upsilon}(t;\lambda)$ hardly changes in time and becomes a nearly time-independent superoperator, which is known as the Markovian limit. In the present calculations, the convergent time $t_\mathrm{c}$ is typically tens of ps and is shorter than $100$ ps. After the convergence, we carry out the ED of the generator $\breve{\mathcal{L}}_\mathrm{c}(\lambda)\equiv\breve{\mathcal{L}}_{\textrm{S}}(\lambda)+\breve{\Upsilon}(t_\mathrm{c};\lambda)$:
\begin{align}
\breve{\mathcal{L}}_\mathrm{c}(\lambda)=\breve{P}(\lambda)\breve{D}(\lambda)\breve{P}^{-1}(\lambda)
\end{align}
where $\breve{P}(\lambda)$ is the matrix composed of the eigenvectors, and $\breve{D}(\lambda)$ is the diagonal matrix constructed from the corresponding eigenvectors. The generalized reduced density matrix $\hat{\rho}^{\rm S}(t;\lambda)$ at any time after $t_\mathrm{c}$ is directly given by
\begin{align}
\kket{\hat{\rho}^{\rm S}(t;\lambda)}=\breve{P}(\lambda)\exp[\breve{D}(\lambda)(t-t_\mathrm{c})]\breve{P}^{-1}(\lambda)\kket{\hat{\rho}^{\rm S}(t_\mathrm{c};\lambda)}.
\end{align}
The method combining RK4 and ED works efficiently for the long-term dynamics calculation over several orders of time. We have confirmed that the results calculated from the combined method are consistent with those obtained only from the RK4 method in the interval $[0, 1\textrm{ ns}]$.

\end{document}